\documentclass[italian,english]{article}
\usepackage[latin9]{inputenc}
\usepackage{color}
\usepackage{graphicx}
\usepackage{amssymb}

\makeatletter

\newcommand{\lyxaddress}[1]{
\par {\raggedright #1
\vspace{1.4em}
\noindent\par}
}

\makeatother

\usepackage{babel}

\begin{document}

\title{\textbf{The Ligo-Ligo cross correlation for the detection of relic
scalar gravitational waves} }

\author{\textbf{Christian Corda}}

\maketitle

\lyxaddress{\begin{center}
International Institute for Theoretical Physics and Advanced Mathematics
Einstein-Galilei, Via Santa Gonda, 14 - 59100 Prato, Italy 
\par\end{center}}

\lyxaddress{\begin{center}
\textit{E-mail address:} \textcolor{blue}{cordac.galilei@gmail.com}
\par\end{center}}
\begin{abstract}
In the earlier work arXiv:0706.3782 we studied the cross correlation
between the Virgo interferometer and the MiniGRAIL resonant sphere
for the detection of relic scalar gravitational waves (SGWs). We have
shown that the overlap reduction function for the cross correlation
between Virgo and the monopole mode of MiniGRAIL is very small, but
a maximum was also found in the correlation at about $2710Hz$, in
the range of the MiniGRAIL sensitivity. 

In this paper the analysis is improved. After carefully reviewing
the response function of interferometers to SGWs directly in the gauge
of the local observer, the result is used to analyze the cross correlation
between the two LIGO interferometers in their advanced configuration
for a potential detection of relic SGWs and to release a lower bound
for the integration time of such a detection. By using a new, frequency-dependent,
overlap reduction function, a comparison with previous low-frequency
approximations is therefore discussed. 

This is the first time that such a cross-correlation is computed in
all the frequency spectrum of relic SGWs. The computation of a better
signal to noise ratio is very important to understand if a scalar
component of GWs is present at high frequency. With the new overlap
reduction function, an order of magnitude is gained concerning the
integration time for detecting the relic SGWs' signal. This could
be very important for a potential detection if advanced projects further
improve their sensitivity.
\end{abstract}

\lyxaddress{PACS numbers: 04.80.Nn, 04.30.Nk, 04.50.+h}

\section{Introduction}

One of the most important goals of interferometric GWs detectors (for
the current status of GWs interferometers see \cite{key-1}) is the
detection of relic GWs that would carry, if detected, a huge amount
of information on the early stages of the Universe evolution (see
the recent review \cite{key-2}, the recent results \cite{key-3}-\cite{key-7}
and references within). 

The mechanism of production of relic GWs is well known. The quantum
fluctuations in the spacetime geometry during the inflationary era
generated relic GWs which would have imprinted tensor perturbations
on the Cosmic Microwave Background Radiation anisotropy. The GWs perturbations
arise from the uncertainty principle and the spectrum of relic GWs
is generated from the adiabatically-amplified zero-point fluctuations
\cite{key-2}-\cite{key-7}. The detection of relic GWs is the only
way to learn about the evolution of the very early universe, up to
the bounds of the Planck epoch and the initial singularity \cite{key-2}-\cite{key-7}.
The potential production of relic GWs has been extended in the framework
of scalar-tensor theories of gravity in \cite{key-8}. 

In earlier work we studied the cross correlation between the Virgo
interferometer and the MiniGRAIL resonant sphere for the detection
of relic SGWs \cite{key-9}. In such a work we have shown that the
overlap reduction function for the cross correlation between Virgo
and the monopole mode of MiniGRAIL is very small, but a maximum was
also found in the correlation at about $2710Hz$, in the range of
the MiniGRAIL sensitivity. 

In this paper the analysis is improved. After carefully reviewing
the response function of interferometers to SGWs directly in the gauge
of the local observer, the result is used to analyze the cross correlation
between the two LIGO interferometers in their advanced configuration
for a potential detection of relic SGWs and to release a lower bound
for the integration time of such a detection. In this way, by using
a new, frequency-dependent, overlap reduction function, a comparison
with previous low-frequency approximations is realized. Note that
this is the first time that such a cross correlation is computed in
all the frequency spectrum of relic SGWs. Then, our computation of
a better signal to noise ratio is very important to understand if
a scalar component of GWs could be present at high frequency.

More, with the new overlap reduction function, an order of magnitude
is gained concerning the integration time for detecting the signal
from relic SGWs. This could be very important for a potential detection
if advanced projects further improve their sensitivity.

It is important to mention the observational constraints to relic
SGWs. Such relic SGWs can be analyzed in terms of the scalar field
$\Phi$ and characterized by a dimensionless spectrum \cite{key-8}
(notice that in this paper natural units are used: $G=1$, $c=1$
and $\hbar=1$):\begin{equation}
\Omega_{sgw}(f)=\frac{1}{\rho_{c}}\frac{d\rho_{sgw}}{d\ln f},\label{eq: spettro}\end{equation}
where \begin{equation}
\rho_{c}\equiv\frac{3H_{0}^{2}}{8G}\label{eq: densita' critica}\end{equation}
is the (actual) critical density energy of the Universe \cite{key-13}.

Concerning an inflationary flat spectrum, recent computations released
the bound \cite{key-8,key-14} \begin{equation}
\Omega_{sgw}(f)h_{100}^{2}<10^{-13}.\label{eq: limite spettro WMAP}\end{equation}

The cross correlation between the two LIGO interferometers which is
computed in the present work is fundamental for the high-frequency
portion of the spectrum (\ref{eq: limite spettro WMAP}) that falls
in the frequency-range of Earth based interferometers. Such a frequency-range
is the interval \cite{key-1} \begin{equation}
10Hz\leq f\leq1KHz.\label{eq: intervallo 1}\end{equation}
Previous cross-correlations, like the one in \cite{key-12}, only
covered the shorter low-frequencies portion 

\begin{equation}
10Hz\leq f\leq100Hz.\label{eq: intervallo vecchio}\end{equation}
In fact, earlier overlap reduction functions were constructed with
approximated low-frequency response functions which did not cover
the high-frequency portion of the interval $100Hz\leq f\leq1KHz.$

\section{A review of the frequency-dependent response function}

The frequency-dependent response function for SGWs has been carefully
computed in \cite{key-9,key-10} where previous computations in the
low frequency approximation \cite{key-11,key-12} have been generalized.

As detectors of GW work in a laboratory environment on Earth \cite{key-1,key-10},
it is quite important to compute the response function in the gauge
of the local observer \cite{key-1}. In this review section we retrieve
the same results of \cite{key-9,key-10}, where the computation was
realized in the transverse-traceless (TT) gauge. 

In the gauge of the local observer, SGWs manifest themselves by exerting
tidal forces on the masses (the mirror and the beam-splitter in the
case of an interferometer). In this gauge the time coordinate $x_{0}$
is the proper time of the observer O, spatial axes are centered in
O and in the special case of zero acceleration and zero rotation the
spatial coordinates $x_{j}$ are the proper distances along the axes
and the frame of the local observer reduces to a local Lorentz frame
\cite{key-16}. In that case the line element reads \cite{key-16}

\begin{equation}
ds^{2}=-(dx^{0})^{2}+\delta_{ij}dx^{i}dx^{j}+O(|x^{j}|^{2})dx^{\alpha}dx^{\beta},\label{eq: metrica local lorentz}\end{equation}
and the effect of GWs on test masses is described by the equation
for geodesic deviation \cite{key-16}

\begin{equation}
\ddot{x^{i}}=-\widetilde{R}_{0k0}^{i}x^{k},\label{eq: deviazione geodetiche}\end{equation}
where $\widetilde{R}_{0k0}^{i}$ are the components of the linearized
Riemann tensor. 

By following \cite{key-9,key-10,key-18}, a good way to analyze variations
in the proper distance (time) is by means of {}``bouncing photons''.
A photon can be launched from the interferometer's beam-splitter to
be bounced back by the mirror. In the gauge of the local observer,
two different effects have to be considered in the calculation of
the variation of the round-trip time for photons. The variations of
the coordinates of the mirror of the interferometer in presence of
a SGW, in the frame of the local observer are \cite{key-10}

\begin{equation}
\delta x(t)=\frac{1}{2}x_{0}\Phi(t)\label{eq: spostamento lungo x}\end{equation}
and

\begin{equation}
\delta y(t)=\frac{1}{2}y_{0}\Phi(t).\label{eq: spostamento lungo y}\end{equation}

$\Phi(t+z)$ is the amplitude of the SGW, $x_{0}$ and $y_{0}$ are
the the unperturbed values of the variables, see \cite{key-10} for
details. Eqs. (\ref{eq: spostamento lungo x}) and (\ref{eq: spostamento lungo y})
have been obtained by using the perturbation method \cite{key-16}
from the equation of motion which are \cite{key-10}

\begin{equation}
\ddot{x}=\frac{1}{2}\ddot{\Phi}x\label{eq: accelerazione mareale lungo x}\end{equation}
and

\begin{equation}
\ddot{y}=\frac{1}{2}\ddot{\Phi}y.\label{eq: accelerazione mareale lungo y}\end{equation}

To compute the response function for an arbitrary propagating direction
of the SGW one recalls that the arms of the interferometer are in
the $\overrightarrow{u}$ and $\overrightarrow{v}$ directions, while
the $x,y,z$ frame is adapted to the propagating SGW. Then, a spatial
rotation of the coordinate system has to be realized

\begin{equation}
\begin{array}{ccc}
u & = & -x\cos\theta\cos\phi+y\sin\phi+z\sin\theta\cos\phi\\
\\v & = & -x\cos\theta\sin\phi-y\cos\phi+z\sin\theta\sin\phi\\
\\w & = & x\sin\theta+z\cos\theta,\end{array}\label{eq: rotazione}\end{equation}
or, in terms of the $x,y,z$ frame

\begin{equation}
\begin{array}{ccc}
x & = & -u\cos\theta\cos\phi-v\cos\theta\sin\phi+w\sin\theta\\
\\y & = & u\sin\phi-v\cos\phi\\
\\z & = & u\sin\theta\cos\phi+v\sin\theta\sin\phi+w\cos\theta.\end{array}\label{eq: rotazione 2}\end{equation}

In this way, the SGW is propagating from an arbitrary direction $\overrightarrow{r}$
to the interferometer (see figure 1). By assuming that the mirror
of eqs. (\ref{eq: spostamento lungo x}) and (\ref{eq: spostamento lungo y})
is located in the $u$ direction, if one uses eqs. (\ref{eq: rotazione}),
(\ref{eq: rotazione 2}), (\ref{eq: spostamento lungo x}) and (\ref{eq: spostamento lungo y}),
the $u$ coordinate of the mirror is 

\begin{figure}

\includegraphics{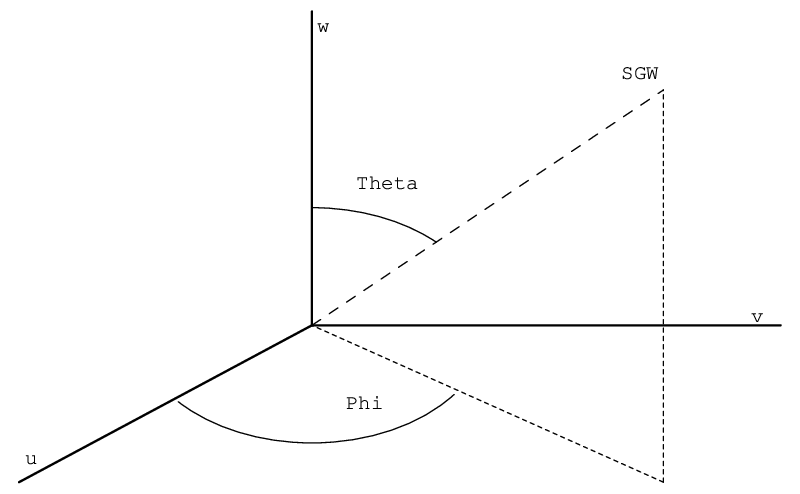}\caption{a SGW propagating from an arbitrary direction, adapted from ref. \cite{key-10}}

\end{figure}

\begin{equation}
u=L+\frac{1}{2}LA\Phi(t-u\sin\theta\cos\phi),\label{eq: du}\end{equation}
where \begin{equation}
A\equiv\cos^{2}\theta\cos^{2}\phi+\sin{}^{2}\phi.\label{eq: A}\end{equation}

We start with a photon which propagates in the $u$ axis. The analysis
is almost the same for a photon which propagates in the $v$ axis.
By putting the origin of the coordinate system in the beam splitter
of the interferometer and using eq. (\ref{eq: du}), the unperturbed
coordinates of the beam-splitter and of the mirror are $u_{b}=0$
and $u_{m}=L$. The unperturbed propagation time between the two masses
is

\begin{equation}
T=L.\label{eq: tempo imperturbato}\end{equation}

From eq. (\ref{eq: du}) the displacements of the two masses under
the influence of the SGW are

\begin{equation}
\delta u_{b}(t)=0\label{eq: spostamento beam-splitter}\end{equation}
and

\begin{equation}
\delta u_{m}(t)=\frac{1}{2}LA\Phi(t-L\sin\theta\cos\phi).\label{eq: spostamento mirror}\end{equation}

The relative displacement, which is defined by

\begin{equation}
\delta L(t)=\delta u_{m}(t)-\delta u_{b}(t),\label{eq: spostamento relativo}\end{equation}
gives

\begin{equation}
\frac{\delta T(t)}{T}=\frac{\delta L(t)}{L}=\frac{1}{2}LA\Phi(t-L\sin\theta\cos\phi).\label{eq: strain magnetico}\end{equation}
But, for a large separation between the test masses (in the case of
Virgo the distance between the beam-splitter and the mirror is three
kilometers, four in the case of LIGO), the definition (\ref{eq: spostamento relativo})
for relative displacements is not physically viable because the two
test masses are taken at the same time and therefore cannot be in
a casual connection \cite{key-10}. The correct definitions for the
bouncing photon are

\begin{equation}
\delta L_{1}(t)=\delta u_{m}(t)-\delta u_{b}(t-T_{1})\label{eq: corretto spostamento B.S. e M.}\end{equation}
and

\begin{equation}
\delta L_{2}(t)=\delta u_{m}(t-T_{2})-\delta u_{b}(t),\label{eq: corretto spostamento B.S. e M. 2}\end{equation}
where $T_{1}$ and $T_{2}$ are the photon propagation times for the
forward and return trip correspondingly. $t$ is the time at which
the photon completes its travel down the arm \cite{key-19}. Thus,
in (21), $t$ is the time that the photon strikes the mirror \cite{key-19}.
In (22), $t$ is the time that the photon strikes the beam splitter
\cite{key-19}. According to the new definitions, the displacement
of one test mass is compared with the displacement of the other at
a later time in order to allow a finite delay for the light propagation
\cite{key-10}. The propagation times $T_{1}$ and $T_{2}$ in eqs.
(\ref{eq: corretto spostamento B.S. e M.}) and (\ref{eq: corretto spostamento B.S. e M. 2})
can be replaced with the nominal value $T$ because the test mass
displacements are already first order in $\Phi$ \cite{key-10}. The
total change in the distance between the beam splitter and the mirror,
in one round-trip of the photon, is

\begin{equation}
\delta L_{r.t.}(t)=\delta L_{1}(t-T)+\delta L_{2}(t)=2\delta u_{m}(t-T)-\delta u_{b}(t)-\delta u_{b}(t-2T),\label{eq: variazione distanza propria}\end{equation}
and in terms of the amplitude of the SGW

\begin{equation}
\delta L_{r.t.}(t)=LA\Phi(t-L\sin\theta\cos\phi-L).\label{eq: variazione distanza propria 2}\end{equation}
In eqs. (23) and (24) $t$ is the time that the photon strikes the
beam splitter \cite{key-19}. The change in distance (\ref{eq: variazione distanza propria 2})
leads to changes in the round-trip time for photons propagating between
the beam-splitter and the mirror

\begin{equation}
\frac{\delta_{1}T(t)}{T}=A\Phi(t-L\sin\theta\cos\phi-L).\label{eq: variazione tempo proprio 1}\end{equation}

In the last calculation (variations in the photon round-trip time
which come from the motion of the test masses inducted by the SGW),
we implicitly assumed that the propagation of the photon between the
beam-splitter and the mirror of the interferometer is uniform as if
it moved in a flat space-time. But, the presence of the tidal forces
indicates that the space-time is curved \cite{key-16}. As a result
one more effect after the first discussed has to be considered, which
requires spacial separation \cite{key-10}. 

From eqs. (\ref{eq: accelerazione mareale lungo x}), (\ref{eq: accelerazione mareale lungo y}),
(\ref{eq: rotazione}) and (\ref{eq: rotazione 2}) the tidal acceleration
of a test mass caused by the SGW in the $u$ direction is\begin{equation}
\ddot{u}(t-u\sin\theta\cos\phi)=\frac{1}{2}LA\ddot{\Phi}(t-u\sin\theta\cos\phi).\label{eq: acc}\end{equation}

Equivalently one can say that there is a gravitational potential \cite{key-10,key-16}

\begin{equation}
V(u,t)=-\frac{1}{2}LA\int_{0}^{u}\ddot{\Phi}(t-l\sin\theta\cos\phi)dl,\label{eq:potenziale in gauge Lorentziana}\end{equation}
which generates the tidal forces, and that the motion of the test
mass is governed by the Newtonian equation

\begin{equation}
\ddot{\overrightarrow{r}}=-\bigtriangledown V.\label{eq: Newtoniana}\end{equation}

For the second effect, the interval for photons propagating along
the $u$ - axis can be written like\begin{equation}
ds^{2}=g_{00}dt^{2}-du^{2}.\label{eq: metrica osservatore locale}\end{equation}

The condition for a null trajectory $ds=0$ gives the coordinate velocity
of the photons 

\begin{equation}
v_{p}^{2}\equiv(\frac{du}{dt})^{2}=1+2V(t,u),\label{eq: velocita' fotone in gauge locale}\end{equation}
which, to first order in $\Phi,$ is approximated by

\begin{equation}
v_{p}\approx\pm[1+V(t,u)],\label{eq: velocita fotone in gauge locale 2}\end{equation}
with $+$ and $-$ for the forward and return trip respectively. By
knowing the coordinate velocity of the photon, the propagation time
for its traveling between the beam-splitter and the mirror can be
defined:

\begin{equation}
T_{1}(t)=\int_{u_{b}(t-T_{1})}^{u_{m}(t)}\frac{du}{v_{p}}\label{eq:  tempo di propagazione andata gauge locale}\end{equation}
and

\begin{equation}
T_{2}(t)=\int_{u_{m}(t-T_{2})}^{u_{b}(t)}\frac{(-du)}{v_{p}}.\label{eq:  tempo di propagazione ritorno gauge locale}\end{equation}

The calculations of these integrals would be complicated because the
$u_{m}$ boundaries of them are changing with time \cite{key-10}

\begin{equation}
u_{b}(t)=0\label{eq: variazione b.s. in gauge locale}\end{equation}
and

\begin{equation}
u_{m}(t)=L+\delta u_{m}(t).\label{eq: variazione specchio nin gauge locale}\end{equation}

But, to first order in $\Phi$, these contributions can be approximated
by $\delta L_{1}(t)$ and $\delta L_{2}(t)$ (see eqs. (\ref{eq: corretto spostamento B.S. e M.})
and (\ref{eq: corretto spostamento B.S. e M. 2})). The combined effect
of the varying boundaries is given by $\delta_{1}T(t)$ in eq. (\ref{eq: variazione tempo proprio 1})
and only the times for photon propagation between the fixed boundaries
$0$ and $L$ have to be computed \cite{key-10}. Such propagation
times will be indicated with $\Delta T_{1,2}$ to distinguish from
$T_{1,2}$. In the forward trip, the propagation time between the
fixed limits is

\begin{equation}
\Delta T_{1}(t)=\int_{0}^{L}\frac{du}{v_{p}(t',u)}\approx L-\int_{0}^{L}V(t',u)du,\label{eq:  tempo di propagazione andata  in gauge locale}\end{equation}
where $t'$ is the delay time (i.e. $t$ is the time at which the
photon arrives in the position $L$, so $L-u=t-t'$) which corresponds
to the unperturbed photon trajectory: 

\begin{center}
$t'=t-(L-u)$. 
\par\end{center}

Similarly, the propagation time in the return trip is

\begin{equation}
\Delta T_{2}(t)=L-\int_{L}^{0}V(t',u)du,\label{eq:  tempo di propagazione andata  in gauge locale 2}\end{equation}
where now the delay time is given by

\begin{center}
$t'=t-u$.
\par\end{center}

The sum of $\Delta T_{1}(t-T)$ and $\Delta T_{2}(t)$ gives the round-trip
time for photons traveling between the fixed boundaries. The deviation
of this round-trip time (distance) from its unperturbed value $2T$
is\begin{equation}
\begin{array}{c}
\delta_{2}T(t)=-\int_{0}^{L}[V(t-2L+u,u)du+\\
\\-\int_{L}^{0}V(t-u,u)]du,\end{array}\label{eq: variazione tempo proprio 2}\end{equation}
and, by using eq. (\ref{eq:potenziale in gauge Lorentziana}), 

\begin{equation}
\begin{array}{c}
\delta_{2}T(t)=\frac{1}{2}LA\int_{0}^{L}[\int_{0}^{u}\ddot{\Phi}(t-2T+l(1-\sin\theta\cos\phi))dl+\\
\\-\int_{0}^{u}\ddot{\Phi}(t-l(1+\sin\theta\cos\phi)dl]du.\end{array}\label{eq: variazione tempo proprio 2 rispetto h}\end{equation}

The total round-trip proper time in presence of the SGW is

\begin{equation}
T_{t}=2T+\delta_{1}T+\delta_{2}T,\label{eq: round-trip  totale in gauge locale}\end{equation}
and\begin{equation}
\delta T_{u}=T_{t}-2T=\delta_{1}T+\delta_{2}T\label{eq:variaz round-trip totale in gauge locale}\end{equation}
is the total variation of the proper time for the round-trip of the
photon in presence of the SGW in the $u$ direction.

By using eqs. (\ref{eq: variazione tempo proprio 1}), (\ref{eq: variazione tempo proprio 2 rispetto h})
and the Fourier transform of $\Phi$ defined by

\begin{equation}
\tilde{\Phi}(\omega)=\int_{-\infty}^{\infty}dt\Phi(t)\exp(i\omega t),\label{eq: trasformata di fourier}\end{equation}
the quantity (\ref{eq:variaz round-trip totale in gauge locale})
can be computed in the frequency domain \cite{key-10} as 

\begin{equation}
\tilde{\delta}T_{u}(\omega)=\tilde{\delta}_{1}T(\omega)+\tilde{\delta}_{2}T(\omega)\label{eq:variaz round-trip totale in gauge locale 2}\end{equation}
where

\begin{equation}
\tilde{\delta}_{1}T(\omega)=\exp[i\omega L(1+\sin\theta\cos\phi)]LA\tilde{\Phi}(\omega)\label{eq: dt 1 omega}\end{equation}

\begin{equation}
\begin{array}{c}
\tilde{\delta}_{2}T(\omega)=-\frac{LA}{2}[\frac{-1+\exp[i\omega L(1+\sin\theta\cos\phi)]-iL\omega(1+\sin\theta\cos\phi)}{(1+\sin\theta\cos\phi)^{2}}+\\
\\+\frac{\exp(2i\omega L)(1-\exp[i\omega L(-1+\sin\theta\cos\phi)]+iL\omega-(1+\sin\theta\cos\phi)}{(-1+\sin\theta\cos\phi)^{2}}]\tilde{\Phi}(\omega).\end{array}\label{eq: dt 2 omega}\end{equation}

In the above computation, the derivative and translation theorems
on the Fourier transform have been used. By using eq. (\ref{eq: A}),
the response function of the $u$ arm of the interferometer to the
SGW is obtained

\begin{equation}
\begin{array}{c}
H_{u}(\omega)=\frac{\tilde{\delta}T_{u}(\omega)}{L\tilde{\Phi}(\omega)}=\frac{1}{2i\omega L}[-1+\exp(2i\omega L)+\\
\\+\sin\theta\cos\phi((1+\exp(2i\omega L)-2\exp i\omega L(1+\sin\theta\cos\phi))].\end{array}\label{eq: risposta u}\end{equation}

Eq. (\ref{eq: risposta u}) is the same result of equation (148) in
 \cite{key-10}, where the computation has been realized in TT gauge.

The computation for the $v$ arm is similar to the one above. We do
not write down redundant computations and we give the direct result 

\begin{equation}
\begin{array}{c}
H_{v}(\omega)=\frac{\tilde{\delta}T_{u}(\omega)}{L\tilde{\Phi}(\omega)}=\frac{1}{2i\omega L}[-1+\exp(2i\omega L)+\\
\\+\sin\theta\sin\phi((1+\exp(2i\omega L)-2\exp i\omega L(1+\sin\theta\sin\phi))],\end{array}\label{eq: risposta v}\end{equation}
which is exactly the result (149) in \cite{key-10}, where the computation
has been made in the TT gauge.

The total response function is given by the difference of the two
response functions of the two arms \cite{key-1,key-9,key-10}\begin{equation}
H_{tot}(\omega)=H_{u}(\omega)-H_{v}(\omega),\label{eq: risposta totale}\end{equation}
and, by using eqs. (\ref{eq: risposta u}) and (\ref{eq: risposta v}),
one gets\begin{equation}
\begin{array}{c}
H_{tot}(\omega)=\frac{\tilde{\delta}T_{tot}(\omega)}{L\tilde{\Phi}(\omega)}=\frac{\sin\theta}{2i\omega L}\{\cos\phi[1+\exp(2i\omega L)-2\exp i\omega L(1+\sin\theta\cos\phi)]+\\
\\-\sin\phi[1+\exp(2i\omega L)-2\exp i\omega L(1+\sin\theta\sin\phi)]\}.\end{array}\label{eq: risposta totale 2}\end{equation}

This equation gives exactly the total response function (150) in \cite{key-10},
where the computation has been realized in the TT gauge. 

Eq. (\ref{eq: risposta totale 2}) is also in perfect agreement with
the low frequencies response function in \cite{key-12}.

\begin{equation}
H_{tot}(\omega\rightarrow0)=-\sin^{2}\theta\cos2\phi.\label{eq: risposta totale approssimata}\end{equation}

\section{The signal to noise ratio in the LIGO-LIGO cross-correlation for
the detection of relic scalar waves}

By considering a stochastic background of relic SGWs, the complex
Fourier amplitude $\tilde{\Phi}$ is treated as a random variable
with zero mean value in a way similar to in the Fourier domain \cite{key-12}.
By assuming that the relic SGWs are isotropic and stationary, the
ensemble average of the product of two Fourier amplitudes can be written
as \cite{key-12}

\begin{equation}
<\tilde{\Phi}^{*}(f,\hat{\Omega})\tilde{\Phi}(f',\hat{\Omega}')>=\delta(f-f')\delta^{2}(\hat{\Omega},\hat{\Omega}')\tilde{S}_{\Phi}(f),\label{eq: media ampiezze}\end{equation}
where $\hat{\Omega}$ is a unit vector specifying the propagation
direction, and by using the explicit definition of the spectrum (\ref{eq: spettro})
\cite{key-12} 

\begin{equation}
\tilde{S}_{\Phi}(f)=\frac{3H_{0}^{2}\Omega_{sgw}(f)}{8\pi^{3}f^{3}}.\label{eq:legame spettro-ampiezza}\end{equation}
The quantity \begin{equation}
\delta^{2}(\hat{\Omega},\hat{\Omega}')\equiv\delta(\phi-\phi^{\prime})\delta(\cos{\theta}-\cos{\theta^{\prime}})\label{eq: Allen Romano}\end{equation}
is the covariant Dirac delta function on the two-sphere, see \cite{key-23}. 

The optimal strategy for a potential detection of a stochastic background
of relic SGWs requires the cross-correlation of at last two detectors
with uncorrelated noises $n_{i}(t),$ $i=1,2$ \cite{key-12}. By
following \cite{key-12}, given the two outputs over a total observation
time $T$,

\begin{equation}
s_{i}(t)=S_{\Phi}^{i}(t)+n_{i}(t),\label{eq: output detector}\end{equation}
a \emph{signal} $S$ can be constructed:

\begin{equation}
S=\int_{-T/2}^{T/2}s_{1}(t)s_{2}(t')Q(t-t'),\label{eq: output detector 2}\end{equation}
where $Q$ is a suitable filter function, usually chosen to optimize
the signal to noise ratio (SNR) \cite{key-12}

\begin{equation}
SNR=<S>/\Delta S.\label{eq: SNR}\end{equation}

In the above equation $\Delta S$ is the variance of $S$. $Q(t-t')$
is a sharply peaked function about values where $t-t'$ is comparable
to the wave travel time between the two detectors (so that unphysical
correlations are ignored) \cite{key-19}. Hence, by assuming that
the observation time is longer than the light travel time between
detectors, we get 

\begin{equation}
<S>=\frac{H_{0}^{2}}{5\pi^{2}}T*Re\left\{ \int_{0}^{\infty}df\frac{\tilde{Q}(f)\Omega_{sgw}(f)\gamma(f)}{f^{3}}\right\} ,\label{eq: Segnale}\end{equation}
in the frequency domain, where $\gamma(f)$ is the so - called overlap
reduction function defined in \cite{key-17} and adapted to scalar
waves in \cite{key-12}. 

For the computation of the variance, one assumes that, in each detector,
the noise is much greater than the strain due to SGWs, obtaining \cite{key-12}
\begin{equation}
\Delta S^{2}\simeq\frac{T}{2}\int_{0}^{\infty}dfP_{1}(|f|)P_{2}(|f|)|\tilde{Q}(f)|^{2},\label{eq: varianza}\end{equation}
where $P_{i}(|f|)$ is the one-sided power spectral density of the
$i$ detector \cite{key-12}.

By introducing the inner product \cite{key-12}

\begin{equation}
(a,b)\equiv Re\left\{ \int_{0}^{\infty}dfa(f)b(f)P_{1}(f)P_{2}(f)\right\} ,\label{eq: prodotto in}\end{equation}
the squared SNR can be rewritten as

\begin{equation}
(SNR)^{2}=2T(\frac{H_{0}^{2}}{5\pi^{2}})^{2}\frac{(\tilde{Q},\frac{\theta(f)\tilde{Q}(f)\Omega_{sgw}(f)\gamma(f)}{f^{3}P_{1}(|f|)P_{2}(|f|)})}{(\tilde{Q},\tilde{Q})},\label{eq: SNR2}\end{equation}
where the Heaviside step function $\theta(f)$ is introduced in order
to insure that the limits of integration are always over positive
frequencies \cite{key-19}. The above ratio is maximal for \cite{key-12}

\begin{equation}
\tilde{Q}=k\frac{\theta(f)\Omega_{sgw}(f)\gamma(f)}{f^{3}P_{1}(|f|)P_{2}(|f|)},\label{eq: proporzionale}\end{equation}
where $k$ is a (real) overall normalization constant \cite{key-23}.
With this optimal choice the signal to noise ratio becomes \begin{equation}
(SNR)=\sqrt{2T}\frac{H_{0}^{2}}{5\pi^{2}}\sqrt{\int_{0}^{\infty}df\frac{\Omega_{sgw}^{2}(f)\gamma^{2}(f)}{f^{6}P_{1}(|f|)P_{2}(|f|)}}.\label{eq: SNR 3}\end{equation}

\section{The generalized overlap reduction function for the advanced LIGO-LIGO
cross-correlation}

The overlap reduction function for SGWs is given by \cite{key-12}

\begin{equation}
\gamma(f)=\frac{15}{4\pi}\int d\hat{\Omega}\exp(2\pi if\hat{\Omega}\cdot\overrightarrow{r}_{12})H_{1}(f)H_{2}(f),\label{eq: ORF}\end{equation}
where $\overrightarrow{r}_{12}$ is the distance between the two detectors
and $H_{i}(f)$ the angular pattern of the $i$ detector ($i=1,2$).
In the literature, the low-frequency approximated angular pattern
(\ref{eq: risposta totale approssimata}) has been used in the computation
of the overlap reduction functions for stochastic backgrounds of relic
SGWs, see \cite{key-12} and references within. Actually, the analysis
can be improved with the aid of the frequency dependent angular pattern
(\ref{eq: risposta totale 2}). By putting the origin of the coordinate
system in the LIGO site in Hanford we write ($\omega=2\pi f$)\begin{equation}
\begin{array}{c}
\gamma(f)=\frac{15}{4\pi}\int\sin\theta d\theta d\phi\exp(2\pi ifX)\frac{\sin\theta}{4\pi ifL}\{\cos\phi[1+\exp(4\pi ifL)-2\exp2\pi ifL(1+\sin\theta\cos\phi)]+\\
\\-\sin\phi[1+\exp(4\pi ifL)-2\exp2\pi ifL(1+\sin\theta\sin\phi)]\}\cdot\\
\\\frac{\sin(\theta-\theta_{1})}{4\pi ifL}\{\cos(\phi-\phi_{1})[1+\exp(4\pi ifL)-2\exp2\pi ifL(1+\sin(\theta-\theta_{1})\cos(\phi-\phi_{1}))]+\\
\\-\sin(\phi-\phi_{1})[1+\exp(4\pi ifL)-2\exp2\pi ifL(1+\sin(\theta-\theta_{1})\sin(\phi-\phi_{1}))]\},\end{array}\label{eq: ORF 2}\end{equation}
where $\theta_{1}=-28.64^{0}$, $\phi_{1}=71.2^{0}$, $L=4Km$ and

\begin{equation}
\begin{array}{c}
X\equiv d\{\cos(\phi_{1}-\phi)(-\sin\theta_{1}+\cos\theta_{1}(\cos\phi_{1}+\sin\phi_{1}))\cdot\\
\\(-\sin\theta+\cos\theta(\cos\phi+\sin\phi))+(\cos\theta_{1}+\sin\theta_{1}(\cos\phi_{1}+\sin\phi_{1}))\\
\\(\cos\theta+\sin\theta(\cos\phi+\sin\phi))-\sin(\phi_{1}+\phi)\end{array}\cdot\label{eq: X}\end{equation}
with $d=2997.9Km.$ The position and orientation of the two LIGO sites
are well known \cite{key-1}. 

Eq. (\ref{eq: ORF 2}) is the overlap reduction function for an isotropic
stochastic background \cite{key-19}. It is a dimensionless function
of frequency $f$ which depends entirely on the relative positions
and orientations of a pair of detectors \cite{key-20} and which is
angle averaged over all directions \cite{key-19}. Further details
on the overlap reduction function can be find in \cite{key-20}.

The computation of the frequency-dependent overlap reduction function
(\ref{eq: ORF 2}) has to be considered together with the one in \cite{key-9}.
In such a work we studied the cross correlation between the Virgo
interferometer and the MiniGRAIL resonant sphere for the detection
of relic scalar gravitational waves (SGWs). It was shown that the
overlap reduction function for the cross correlation between Virgo
and the monopole mode of MiniGRAIL is very small, but a maximum was
also found in the correlation at about $2710Hz$, in the range of
the MiniGRAIL sensitivity. 

In figure 2, the absolute value of the overlap reduction function
(\ref{eq: ORF 2}) is drawn in the frequency-range of Earth based
interferometers that is the interval (\ref{eq: intervallo 1}). We
emphasize that old overlap reduction functions, constructed with the
approximated angular pattern (\ref{eq: risposta totale approssimata}),
guarantee only the coverage of the shorter interval (\ref{eq: intervallo vecchio}).
The new overlap reduction function of equation (\ref{eq: ORF 2})
guarantees also the coverage of the high frequency portion $100Hz\leq f\leq1KHz.$
This is very important for the potential detection of relic SGW at
high frequencies.

\begin{figure}
\includegraphics{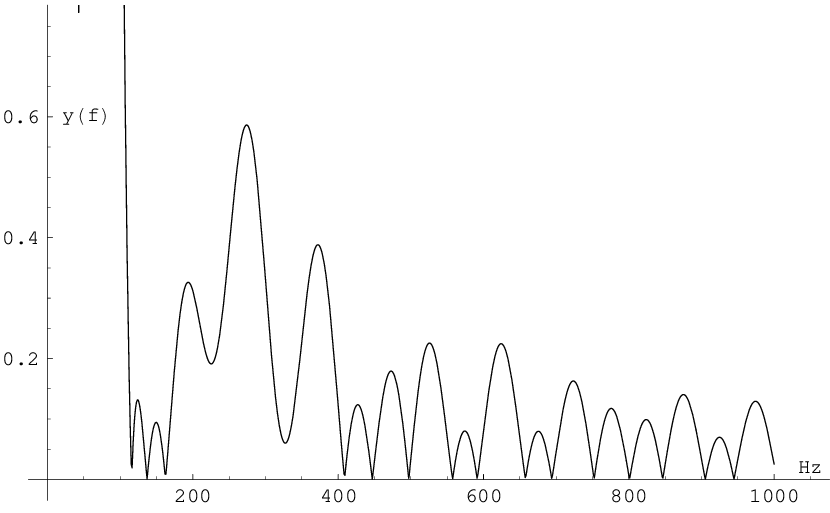}

\caption{The absolute value of the overlap reduction function (\ref{eq: ORF 2})
in the frequency-range $10Hz\leq f\leq1KHz$. }

\end{figure}
In figure 2 the value of the overlap reduction function of the two
LIGO interferometers for SGWs appears low, thus, in principle, we
need a long integration time in order to improve the SNR. A lower
bound for the integration time of a potential detection will be released
in next Section. This lower bound for the integration time corresponds
to an upper bound for the specific SNR of the cross-correlation between
the two advanced LIGO.

\section{The specific signal to noise ratio for scalar waves in a cross-correlation
between the two advanced LIGO interferometers }

The spectrum of relic GWs is flat in the frequency-range of Earth
based interferometers. WMAP observations put strongly severe restrictions
on such a spectrum \cite{key-8,key-13,key-14}. In fig. 3 we map the
spectrum $\Omega_{sgw}$ choosing the amplitude (determined by the
ratio $\frac{M_{inl}}{M_{Planck}}$ \cite{key-8,key-14} to be \textit{as
large as possible, consistent with the WMAP constraints }\cite{key-8,key-13,key-14}.
The inflationary spectrum rises quickly at low frequencies (waves
which re-entered in the Hubble sphere after the Universe became matter
dominated) and falls off above the (appropriately redshifted) frequency
scale $f_{max}$ associated with the fastest characteristic time of
the phase transition at the end of inflation. The amplitude of the
flat region depends on the energy during the inflationary stage \cite{key-8,key-14}.
As WMAP data are consistent with a maximum inflationary scale $M_{inl}=2\cdot10^{16}GeV$
\cite{key-8,key-13,key-14}, this means that today, at LIGO and LISA
frequencies, indicate by the lines in fig. 3, one obtains the bound
of equation (\ref{eq: limite spettro WMAP}).

Let us apply the new overlap reduction function obtained in eq. (\ref{eq: ORF 2}).
To detect a stochastic background of relic scalar GWs we need to obtain
a signal to noise ratio equal, at least, to the one in eq. (\ref{eq: SNR 3}).
By iserting $(SNR)=1$ in eq. (\ref{eq: SNR 3}) and by solving in
respect to the observation time one gets

\begin{equation}
T=(\frac{5\pi^{2}}{H_{0}^{2}})^{2}\frac{1}{2\int_{0}^{\infty}df\frac{\Omega_{sgw}^{2}(f)\gamma^{2}(f)}{f^{6}P_{1}(|f|)P_{2}(|f|)}}.\label{eq: tempo d'osservazione}\end{equation}

Let us insert in eq. (\ref{eq: tempo d'osservazione}) the value of
eq. (\ref{eq: ORF 2}) of the new overlap reduction function and the
value of the spectrum in eq. (\ref{eq: limite spettro WMAP}). For
both of $P_{1}(|f|)\mbox{ }\mbox{ and }P_{2}(|f|)$ we use the analytical
fit of \cite{key-12} for the noise spectral density of advanced LIGO,
which is 

\begin{equation}
P(f)=\frac{P_{0}}{5}\left[(\frac{f_{0}}{f})^{4}+2+2(\frac{f}{f_{0}})^{2}\right]\label{eq: NSD}\end{equation}
where $P_{0}=2.3\cdot10^{-48}Hz^{-1}$ and $f_{0}=75Hz.$

In this way, by integrating the right hand side of eq. (\ref{eq: tempo d'osservazione})
in all the range of eq. (\ref{eq: intervallo 1}), which is the total
frequency range of Earth based interferometers, one obtains 

\begin{equation}
T\sim8*10^{5}years.\label{eq: tempo d'0sservazione 2}\end{equation}

The assumption that all the scalar perturbation in the Universe are
due to a stochastic background of relic SGWs is quit strong, but the
result can be considered like a lower bound for the observation time.

To better understand the difference between the response function
of this paper and previous low-frequency approximated ones a brief
comparison is now discussed. In the previous literature, people inserted
the approximated response function of eq. (\ref{eq: risposta totale approssimata})
in eq. (\ref{eq: tempo d'osservazione}). More, the integration was
performed only in the low frequencies range of eq. (\ref{eq: intervallo vecchio})
where the low frequencies approximation of eq. (\ref{eq: risposta totale approssimata})
is correct.  If one inserts the old, approximated, overlap reduction
function in in eq. (\ref{eq: tempo d'osservazione}) and uses the
same eq. (\ref{eq: NSD}) for both of $P_{1}(|f|)\mbox{ }\mbox{ and }P_{2}(|f|)$
and the same value of the spectrum in eq. (\ref{eq: limite spettro WMAP}),
the observation time obtained from eq. (\ref{eq: tempo d'osservazione})
results overpriced \begin{equation}
T\sim7*10^{6}years.\label{eq: tempo d'osservazione 3}\end{equation}
 An order of magnitude is gained by introducing the new overlap reduction
function and by integrating in all the frequency range of Earth based
interferometers that is the interval of eq. (\ref{eq: intervallo 1}).

The times involved in the two results (the correct one and the overpriced
one) are substantially longer than the entire existence of the human
species \cite{key-19}. In order to be interesting, one has to determine
the necessary sensitivity to bring the observation time down to a
reasonable value (e.g. the lifetime of a graduate student) \cite{key-19}.
For this goal, one notes from eq. (\ref{eq: SNR 3}) that a gain in
sensitivity higher than 2 orders of magnitude, which will permit to
reduce of 5 orders of magnitude the noise spectral density of interferometric
GW detectors, could in principle provide $(SNR)=1$ for observation
times which are not too long (order of some years). In this case,
the order of magnitude gained with the new overlap reduction function
will be very important for the signal detection. This sensitivity
looks to be not beyond the scope of any known technology. In fact,
a fundamental obstacle which limited the sensitivity of interferometric
GW detectors was the vacuum (zero-point) fluctuations of the electromagnetic
field. Recently, it has been shown that a quantum technology\textemdash{}the
injection of squeezed light\textemdash{}offers a solution to this
problem \cite{key-21}. On the other hand, the third generation of
GW detectors will limit the effect of the seismic noise, and, through
cryogenic facilities, will cool down the mirrors to directly reduce
the thermal vibration of the test masses \cite{key-22}. By using
eq. (\ref{eq: NSD}) one gets $P(100\mbox{ }Hz)\sim10^{-47}Hz^{-1}$
which gives a sensitivity in strain of $h\sim10^{-25}$ at $100\mbox{ }Hz,$
i.e. the frequency where the sensitivity of interferometric GW detectors
is highest. The Einstein Telescope will be sensitive to intrinsic
GW amplitudes of order $h\sim10^{-27}$ in the frequency range 6 Hz
to 3 kHz \cite{key-22}. Therefore, a gain in sensitivity higher than
2 orders of magnitude looks a concrete possibility.

\section{Conclusions}

In earlier work we studied the cross correlation between the Virgo
interferometer and the MiniGRAIL resonant sphere for the detection
of relic SGWs. It was shown that the overlap reduction function for
the cross correlation between Virgo and the monopole mode of MiniGRAIL
is very small, but a maximum was also found in the correlation at
about $2710Hz$, in the range of the MiniGRAIL sensitivity. 

In this paper the analysis has been improved. After carefully reviewing
the response function of interferometers to SGWs directly in the gauge
of the local observer, the result has been used to analyze the cross-correlation
between the two LIGO interferometers in their advanced configuration
for a potential detection of relic SGWs and to release a lower bound
for the integration time of such a detection. By using a new, frequency-dependent,
overlap reduction function, a comparison with previous low-frequency
approximations has been performed too. This is the first time that
such a cross-correlation is computed in all the frequency spectrum
of relic SGWs. The computation of a better signal to noise ratio is
very important to understand if a scalar component of GWs is present
at high frequency.

With the new overlap reduction function, an order of magnitude is
gained concerning the integration time for detecting relic SGWs signal,
and this could be very important for a potential detection if advanced
projects further improve their sensitivity.

\begin{figure}
\includegraphics{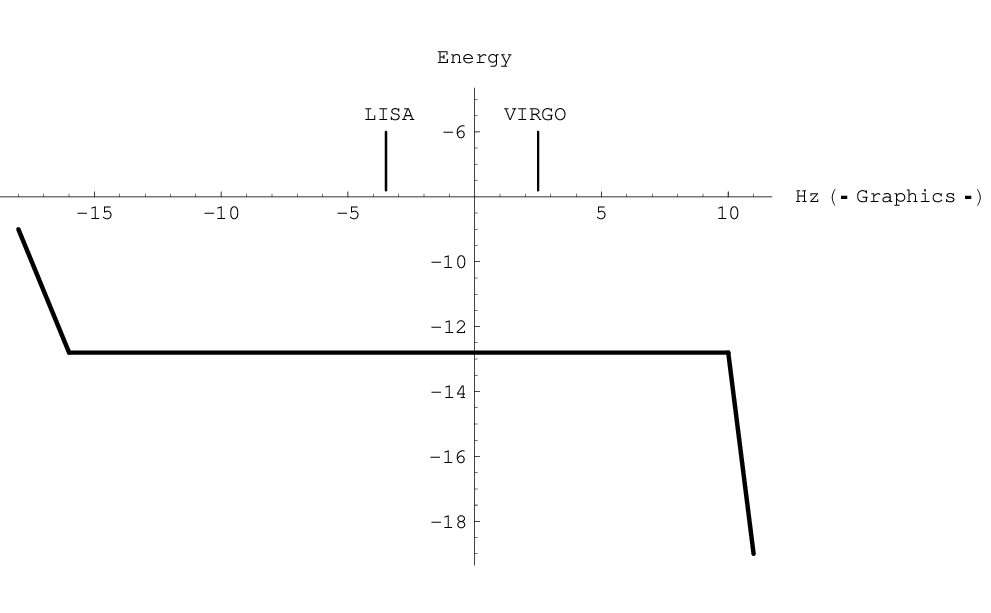}

Figure 3: The spectrum of relic SGWs in inflationary models is flat
over a wide range of frequencies. The horizontal axis is $\log_{10}$
of frequency, in Hz. The vertical axis is $\log_{10}\Omega_{gsw}$.
The inflationary spectrum rises quickly at low frequencies (waves
which re-entered in the Hubble sphere after the Universe became matter
dominated) and falls off above the (appropriately redshifted) frequency
scale $f_{max}$ associated with the fastest characteristic time of
the phase transition at the end of inflation. The amplitude of the
flat region depends on the energy during the inflationary stage; we
have chosen the largest amplitude consistent with the WMAP constrains
on scalar perturbations. This means that at LIGO and LISA frequencies,
$\Omega_{sgw}(f)h_{100}^{2}<9*10^{-13}$. Adapted from ref. \cite{key-8}

\end{figure}

\section{Acknowledgements}

The R. M. Santilli Foundation has to be thanked for partially supporting
this paper\emph{ }(Research Grant Number RMS-TH-5735A2310\emph{).
}I thank an unknown referee for useful comments. 
\selectlanguage{italian}%

\end{document}